\documentclass[aps,reprint,superscriptaddress,preprintnumbers,longbibliography,nofootinbib]{revtex4-1}

\usepackage{amsthm}
\usepackage{graphicx}
\usepackage{amsmath}
\usepackage{hyperref}
\usepackage{xspace}
\usepackage{braket}
\usepackage{todonotes}
\usepackage{subcaption}

\usepackage{dcolumn}
\usepackage{bm}

\newtheorem*{theorem*}{Theorem}

\DeclareRobustCommand{\Sec}[1]{Sec.~\ref{#1}}

\DeclareRobustCommand{\App}[1]{App.~\ref{#1}}

\DeclareRobustCommand{\Fig}[1]{Fig.~\ref{#1}}

\DeclareRobustCommand{\Eq}[1]{Eq.\,(\ref{#1})}
\DeclareRobustCommand{\Eqs}[2]{Eqs.\,(\ref{#1}) and (\ref{#2})}
\DeclareRobustCommand{\Ref}[1]{Ref.\,\cite{#1}}
\DeclareRobustCommand{\Refs}[1]{Refs.\,\cite{#1}}

\usepackage{color}
\definecolor{darkblue}{rgb}{0,0,0.5}
\definecolor{darkred}{rgb}{0.5,0,0}
\definecolor{darkgreen}{rgb}{0,0.5,0}

\newcommand{\be}{\begin{equation}}
\newcommand{\ee}{\end{equation}}

\allowdisplaybreaks

\begin{document}

\preprint{SLAC-PUB-17645}

\title{A Challenge for Discrimination of Color-Singlet\\versus Color-Octet Quarkonium Production}

\author{Andrew J.~Larkoski}
\email{larkoski@slac.stanford.edu}
\affiliation{SLAC National Accelerator Laboratory, Menlo Park, CA 94025, USA}

\date{\today}

\begin{abstract}
The precise mechanism for production of quarkonium at hadron colliders is still an open question. Within non-relativistic quantum chromodynamics, quarkonium production cross sections can be factorized into short-distance, perturbative contributions and universal, non-perturbative, long-distance matrix elements, and then summed over quantum numbers of the heavy quark pair.  In principle, at short-distances, the heavy quark pair can be either in a color-singlet or color-octet state, and it is desirable to establish the relative contributions to compare with data and to make predictions in different experimental environments.  From the explicit form of the lowest-order perturbative matrix elements for color-singlet and color-octet production, we show that the structure of the optimal observable for discrimination on phase space, the likelihood ratio, has strong dependence on the angular momentum state of the heavy quark pair.  This presents an obstruction for construction of a single, robust discrimination observable that can be applied to production of an arbitrary quarkonium state.
\end{abstract}

\maketitle

\tableofcontents

\section{Introduction}

The observation and experimental analysis of bound states of heavy quarks is a fruitful system for strong tests of quantum chromodynamics (QCD) (for reviews, see, e.g., \Refs{Kramer:2001hh,QuarkoniumWorkingGroup:2004kpm,Lansberg:2008gk,Brambilla:2010cs,Bodwin:2013nua,Andronic:2015wma,Lansberg:2019adr}).  At a hadron collider, like the Large Hadron Collider, the production mechanism for these quarkonia can be described within low-energy models or effective theories of QCD.  In this paper, we will focus on the structure of predictions from non-relativistic QCD (NRQCD) \cite{Bodwin:1994jh}, for which a differential cross section $d\sigma$ for quarkonium production can be expressed at leading-twist in factorized, schematic form as
\begin{align}
d\sigma = \sum_n d\hat \sigma_n\, \langle {\cal O}_n\rangle\,.
\end{align}
Here, $d\hat \sigma_n$ is the short-distance, partonic, perturbative QCD cross section and $\langle {\cal O}_n\rangle$ is a universal non-perturbative, long-distance matrix element (LDME), that describes the fragmentation of the perturbative heavy quark--anti-quark pair into quarkonium.  The sum in $n$ runs over the different heavy quark states whose quantum numbers, such as total angular momentum, are consistent with the final quarkonium of interest.  The expansion parameters of NRQCD are both the strong coupling $\alpha_s$ in the perturbative cross section and the speed $v$ of the heavy quarks about their center-of-mass.  For predictive power, the sum over heavy quark pair states is truncated at a fixed order in $\alpha_s$ and $v$, and only a small number of LDMEs need to be extracted from data.

Hadrons are of course singlets under the color symmetry of QCD, but within NRQCD, the perturbative heavy quark states can be either singlets or octets under color.  The necessity of including the color-octet states in predictions of inclusive production at hadron colliders has been borne out in extensive experimental analyses, e.g., \Refs{CDF:2000pfk,CDF:2007msx,LHCb:2011zfl,LHCb:2012geo,LHCb:2013itw,LHCb:2013izl,LHCb:2014brf,LHCb:2019eaj,ATLAS:2011aqv,ATLAS:2014ala,ATLAS:2017prf,CMS:2011rxs,CMS:2013gbz,CMS:2015lbl,CMS:2019jas,ALICE:2011gej,ALICE:2012vup,ALICE:2012vpz,ALICE:2014uja,ALICE:2018crw,ALICE:2019pid}.  However, more exclusive measurements, like of the polarization of the quarkonium as determined from its decay to leptons, suggest some tension or disagreement between experimental data and NRQCD predictions.  Further, the factorization imposed in NRQCD for initial hadronic states has never conclusively been proven, and has challenges for experimental validation of universality of the LDMEs \cite{Rothstein:1999vz}.  More concrete and predictive formulations of NRQCD have been proposed since, e.g., \Refs{Luke:1999kz,Brambilla:2004jw,Nayak:2005rt}, but these issues mean that within NRQCD the production mechanism for quarkonia at hadron colliders is still an open question.

Regardless of the validity of NRQCD, we can still use its factorization as a guide for testing the production mechanism of quarkonia.  In this paper, we study observables for distinguishing color-singlet versus color-octet production at short distances in a hadron collider.  We will assume that we only measure particle momentum, and so the particular angular momentum states of the heavy quark pair is unknown.  Therefore, we can only make predictions that are inclusive over all angular momentum states, consistent with the quantum numbers of the observed quarkonium.  Distinguishing the color-octet from color-singlet production is a binary discrimination problem, for which the optimal observable is the likelihood ratio, by the Neyman-Pearson lemma \cite{Neyman:1933wgr}.  As a physical observable that we could perform cuts on to purify a quarkonium sample, the likelihood ratio is then dependent on the perturbative, partonic cross section, as well as the LDMEs:
\begin{align}
{\cal L} =\frac{d\sigma^{(8)}}{d\sigma^{(1)}}= \frac{\sum_n d\hat \sigma_n^{(8)}\langle {\cal O}_n^{(8)}\rangle}{\sum_n d\hat \sigma_n^{(1)}\langle {\cal O}_n^{(1)}\rangle}\,,
\end{align}
where here $n$ represents the possible angular momentum states of the heavy quark pair.  The exponents $(8)$ or $(1)$ represent the color-octet or color-singlet states, respectively.  This particular form for the likelihood ratio isn't directly useful for predictions because its functional form will depend in detail on the functional dependence of momenta in the short-distance cross sections $d\hat \sigma_n$, as well as on the particular values of the LDMEs, $\langle {\cal O}_n\rangle$.  However, the form of this likelihood ratio as calculated analytically can be used as inspiration for a simple, robust observable that can be used in data to discriminate these production mechanisms.  Such an observable will perhaps not be the optimal observable for discrimination, but can be powerful, nonetheless.

We will show that this proposed observable construction procedure is undermined by the analytic form of the short-distance amplitudes.  In particular, we show that the likelihood ratios for individual angular momentum states as calculated from lowest-order perturbative amplitudes have non-trivial kinematic dependence on Mandelstam invariants $s,t,u$ that changes with the angular momentum of the heavy quark pair.\footnote{For a specific quarkonium of interest, only some of the angular momentum and color states contribute to fixed-order in the velocity expansion of NRQCD.  So, the likelihood ratio of color-octet to color-singlet matrix elements with the same angular momentum quantum numbers does not directly apply to NRQCD predictions for specific hadrons, like $J/\psi$ production.  However, our goal here is applicability to general quarkonia and considerations beyond strict NRQCD factorization assumptions.}  For angular momentum states $^1S_0$, $^3S_1$, and $^3P_0$, their likelihood ratios take minimum and maximum values at the same points on $s,t,u$ phase space.  This suggests that a single, common observable can be used to efficiently discriminate color-singlet and -octet production of these states.  However, the maximum and minimum values of the likelihood ratios for the states $^3P_1$ and $^3P_2$ occur at the exact opposite points on $s,t,u$ phase space as for states $^1S_0$, $^3S_1$, and $^3P_0$.  This implies that a universal discrimination observable applicable to production of any quarkonium state cannot be constructed from simply interpolating between phase space points of maximum and minimum perturbative likelihood.  Because of this, knowledge about the specific quarkonium state produced and the matrix elements that contribute is required to distinguish color-singlet versus color-octet production.

This paper is organized as follows. In \Sec{sec:notkin}, we introduce the notation we employ throughout this paper, and present the expressions for the kinematics we consider.  In \Sec{sec:discstates}, we present the lowest-order perturbative amplitudes for color-singlet and color-octet quarkonium production, and their corresponding likelihood ratios, for each angular momentum state.  In this section, we also identify simple, robust observables for discrimination that can interpolate between extrema of the likelihood ratios.  We summarize our results and conclude in \Sec{sec:conc}.  Explicit formulas for the $^3P_J$ amplitudes are presented in an appendix.   In another appendix, we also present comparisons of color-singlet and color-octet matrix elements that contribute to the production of $\chi_J$ hadrons at leading power in NRQCD and show that their structure depends on the total angular momentum $J$.

\section{Notation and Kinematics}\label{sec:notkin}

In this paper, we will study features of the amplitudes for production of a heavy quark pair denoted as $[Q\bar Q]$ at a hadron collider.  This heavy quark pair then subsequently fragments into a quarkonium hadron.  We will restrict our analysis to the lowest-order short-distance process that allows for non-trivial kinematics of the quarkonium with exclusively strong interactions, namely $gg\to [Q\bar Q]g$.  Of course, there are also non-zero amplitudes for production processes involving light quarks, like $q\bar q \to [Q\bar Q]g$, but we do not consider their effect here.  The reason for this is that lowest-order production processes involving quarks do not display any sensitivity to the color representation of the produced heavy quark pair, as can be explicitly demonstrated by comparing amplitudes from \Ref{Baier:1983va} and \Ref{Cho:1995ce}.  Actually, this insensitivity to the color of the heavy quark pair introduces another challenge to distinguish color-singlet from color-octet production, beyond what we will establish here.  We leave a thorough analysis of realistic quarkonium production, including parton distributions, LDMEs, detector acceptance, etc., to future work.  Such a more detailed analysis could be performed with numerical programs for quarkonia production, like HELAC-Onia \cite{Shao:2012iz,Shao:2015vga} or Mad-Onia \cite{Artoisenet:2007qm}, but we will restrict to analytic amplitudes for more direct and unambiguous functional comparison.

With these limitations established, the short-distance, perturbative amplitudes we consider take the form
\begin{align}
|{\cal A}(g_1g_2\to [Q\bar Q] g_3)|^2 \equiv f(s,t,u)\,,
\end{align}
where the Mandelstam invariants are
\begin{align}
s = (p_1+p_2)^2\,, \ \ t = (p_1-p_3)^2\,, \ \ u = (p_2-p_3)^2\,,
\end{align}
in terms of the momenta of the three gluons in the process.  We also denote the mass of quarkonium as $m^2_{Q\bar Q} \equiv M^2 = s+t+u$, by momentum conservation and the masslessness of the gluons.  By Bose symmetry, the functional form of the amplitude must be completely permutation-symmetric in $s,t,u$:
\begin{align}
f(s,t,u)=f(u,s,t) = f(t,u,s)\,.
\end{align}
For quarkonium production, the corresponding phase space can be defined on the $(s,t)$ plane with the constraints $s+t\geq M^2$, $s\geq M^2$, $t\leq 0$.  This is semi-infinite in both $s$ and $t$ and so a compact way to represent the phase space is to cross the gluons to the final state and the heavy quark pair to the initial state to describe quarkonium decay to gluons: $[Q\bar Q]\to ggg$.  Then, all of the Mandelstams $s,t,u\geq 0$ and the phase space is the triangular region bounded above by $s+t\leq M^2$.  We will exclusively display plots on this quarkonium decay phase space for compactness, but will consider some results for production explicitly.  Of course, the expressions for the squared amplitudes describing production or decay are identical by crossing symmetry, but their interpretation in a particular frame will depend on the process.

The particular heavy quark pair state will be labeled by angular momentum and SU(3) color quantum numbers as $[Q\bar Q]\equiv \, ^{2S+1}L_J^{(D)}$, where $L$ is the orbital angular momentum of the quark and anti-quark system, $S$ is the sum of their spins, $J$ is their total angular momentum, and $D$ is the dimension of their SU(3) color representation.  Because quarks carry color in the fundamental representation of SU(3), the only possible irreducible color states from their product is the adjoint octet ${\bf 8}$ or the singlet ${\bf 1}$.  We will therefore focus on discrimination of the squared amplitudes describing these two color representations of the heavy quark pair, for each angular momentum state $^{2S+1}L_J$.  The particular angular momentum states present and their relative probabilities depend on the LDMEs and experimental cuts, to which we wish to remain agnostic.

For a fixed angular momentum state, the likelihood ratio ${\cal L}$ of interest is defined to be
\begin{align}
{\cal L}\equiv \frac{|{\cal A}(gg\to \,\! ^{2S+1}L_J^{(8)} g)|^2}{|{\cal A}(gg\to \,\! ^{2S+1}L_J^{(1)} g)|^2}\,,
\end{align}
the ratio of color-octet to color-singlet matrix elements.  The limit ${\cal L}\to 0$ selects color-singlet events, while the limit ${\cal L}\to\infty$ selects color-octet events.  Any monotonic function of the likelihood is equivalent in discrimination power, so we can be ignorant of overall multiplicative factors in the amplitudes, and just focus on the kinematic dependence of the likelihood on $s,t,u$.  So, the likelihood will be some function of $s,t,u$; ${\cal L}\equiv {\cal L}(s,t,u)$.  In the expressions of the explicit amplitudes we will consider later, we will ignore all coupling factors, overall constants, and quarkonium wavefunction dependence.  To the reader interested in these factors we point them to the relevant references.

\section{Discrimination of Color States}\label{sec:discstates}

We now present the short-distance amplitudes for heavy quark pair production and the likelihood ratios between the color-octet and -singlet states for each set of angular momentum quantum numbers.  We start with the $S$-wave heavy quark states, and then construct simple, robust infrared and collinear safe observables that interpolate between the extrema of the $S$-wave likelihoods.  This is sensible for $S$-wave states, because the minima and maxima of the likelihoods are located at the same points on phase space.  However, when considering $P$-wave states, we will show that the maxima and minima of the likelihood can lie at different points for different spin configurations, hence complicating the construction of a single, general observable that can provide robust discrimination between color states for all angular momentum states.

In this section, we will explicitly write the squared amplitudes of all $S$-wave heavy quark states as they are all relatively compact.  Squared amplitudes for $P$-wave states are presented in \App{app:me}.

\subsection{$S$-wave States}

\subsubsection{$^1S_0$}

\begin{figure}[t]
\begin{center}
\includegraphics[width=0.5\textwidth]{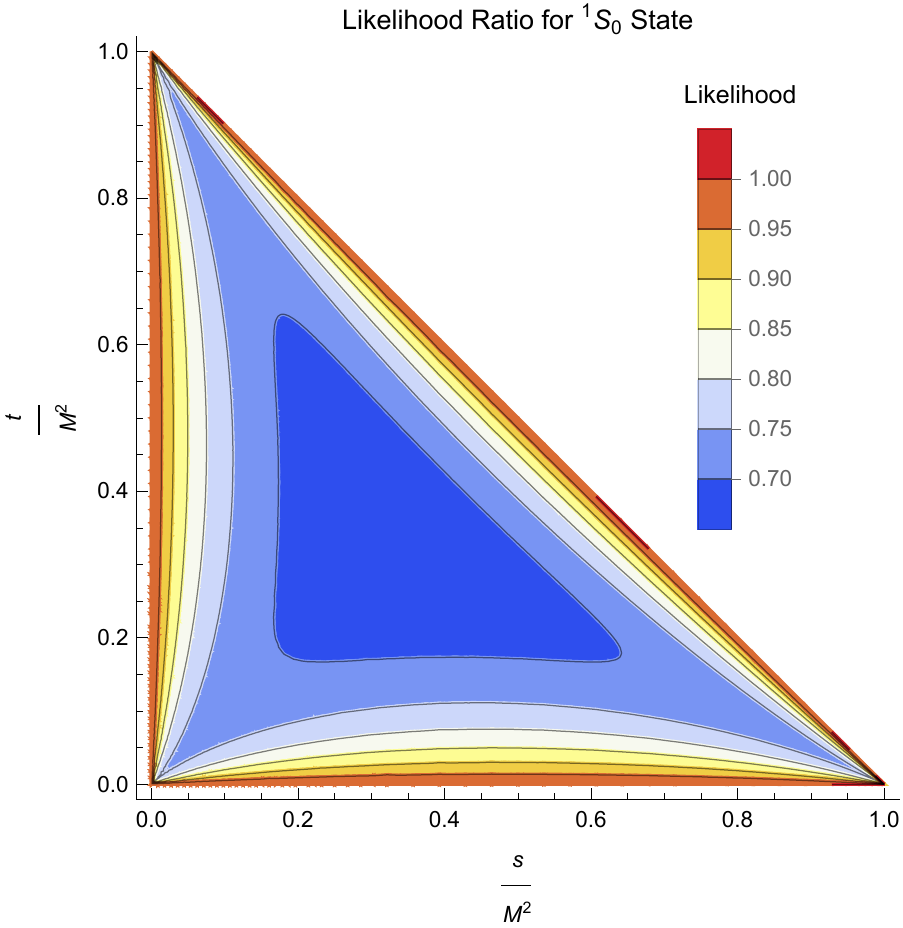}
\caption{\label{fig:like1s0}
Plot of the perturbative octet vs.~singlet likelihood ratio on $(s,t)$ phase space for the $^1S_0$ state.}
\end{center}
\end{figure}

The lowest-order perturbative amplitude for production of the 0 angular momentum, color-singlet heavy quark pair state is \cite{Baier:1983va,Gastmans:1987be}
\begin{align}
&|{\cal A}(gg\to \,\! ^1S_0^{(1)}g)|^2\\
&
\hspace{1cm}\propto \frac{(st+tu+us)^2}{(s+t)^2(t+u)^2(u+s)^2}\frac{M^8+s^4+t^4+u^4}{stu}\,.\nonumber
\end{align}
The corresponding color-octet amplitude is \cite{Cho:1995vh,Cho:1995ce}
\begin{align}
&|{\cal A}(gg\to \,\! ^1S_0^{(8)}g)|^2\\
&
\hspace{1cm}\propto \frac{(st+tu+us)^2-M^2 s tu}{(s+t)^2(t+u)^2(u+s)^2}\frac{M^8+s^4+t^4+u^4}{stu}\,.\nonumber
\end{align}
Their likelihood ratio is then
\begin{align}\label{eq:like1s0}
{\cal L} = 1-\frac{M^2stu}{(st+tu+us)^2}\,.
\end{align}
A contour plot of this on the $(s,t)$ phase space is shown in \Fig{fig:like1s0}.  Note that the minimum value of the likelihood occurs in the region where $s\sim t\sim u\sim M^2/3$, and the maxima lie along the boundary of phase space, where one or two of the Mandelstam invariants vanish.

Specific values of the likelihood ratio in some limits are interesting to identify, and illustrate its range.  There are three limits we will consider: the collinear limit (one Mandelstam invariant vanishes), the soft gluon limit (two Mandelstams vanish), and the hard limit (all Mandelstams are equal to $M^2/3$).  By Bose symmetry, we can study the collinear limit for $u\to 0$, for which the likelihood becomes unity:
\begin{align}
\lim_{u\to 0}{\cal L} = 1\,.
\end{align}
Along the edges of phase space, away from the vertices, the likelihood becomes unity.  Again, by Bose symmetry, the soft limit can be isolated by taking $t,u\to 0$ and $s\to M^2$, but $t\sim u$, for which the likelihood ratio becomes
\begin{align}
\lim_{t,u\to 0}{\cal L} = 1-\frac{tu}{(t+u)^2}\in\left[\frac{3}{4},1\right]\,.
\end{align}
At the vertices of phase space, the value of the likelihood depends on the direction of the soft gluon with respect to the other particles in the process.  Finally, in the hard limit at the center of phase space, the likelihood reduces to
\begin{align}
\lim_{s=t=u}{\cal L} = \frac{2}{3}\,,
\end{align}
which is also where it takes its minimum value.

\subsubsection{$^3S_1$}

\begin{figure}[t]
\begin{center}
\includegraphics[width=0.5\textwidth]{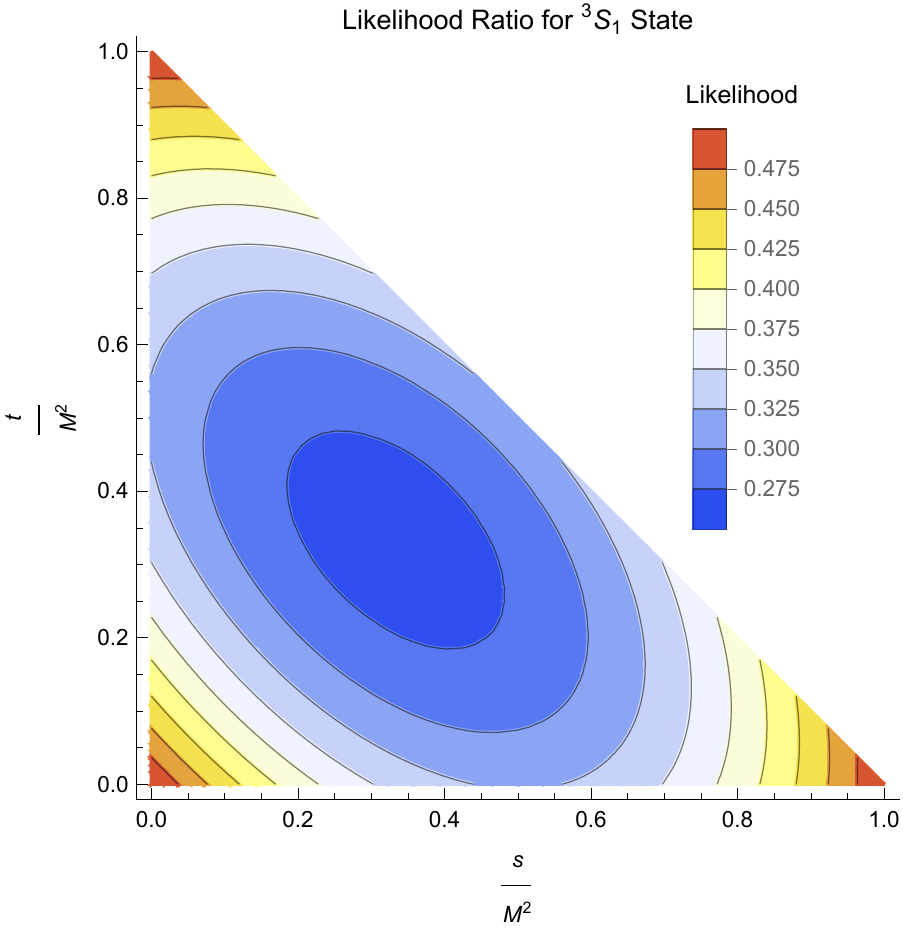}
\caption{\label{fig:like3s1}
Plot of the perturbative octet vs.~singlet likelihood ratio on $(s,t)$ phase space for the $^3S_1$ state.}
\end{center}
\end{figure}

The lowest-order perturbative amplitude for production of the spin-1, $S$-wave color-singlet heavy quark pair state is \cite{Baier:1983va,Gastmans:1987be}
\begin{align}
&|{\cal A}(gg\to\, ^3S_1^{(1)}g)|^2 \\
&
\hspace{1cm}\propto M^2\frac{s^2(t+u)^2+t^2(s+u)^2+u^2(s+t)^2}{(s+t)^2(s+u)^2(t+u)^2}\,.\nonumber
\end{align}
The corresponding color-octet amplitude, when also summed over constituent quark helicities, is \cite{Cho:1995vh,Cho:1995ce}
\begin{align}
&|{\cal A}(gg\to \, ^3S_1^{(8)}g)|^2 \\
&\hspace{0.5cm}\propto \frac{\left(\frac{s^2t^2+t^2u^2+u^2s^2}{M^2}+stu\right)\left(M^4-\frac{27}{19}(st+tu+us)\right)}{(s+t)^2(t+u)^2(u+s)^2}\,.\nonumber
\end{align}
The likelihood ratio is then
\begin{align}\label{eq:like3s1}
{\cal L} = \frac{\left(\frac{s^2t^2+t^2u^2+u^2s^2}{M^2}+stu\right)\left(M^4-\frac{27}{19}(st+tu+us)\right)}{M^2(s^2(t+u)^2+t^2(u+s)^2+u^2(s+t)^2)}\,.
\end{align}
A contour plot of this on the $(s,t)$ phase space is shown in \Fig{fig:like3s1}.  Just like for the $^1S_0$ states, the minimum value of the likelihood occurs in the region where $s\sim t\sim u\sim M^2/3$, and the maxima lie at the vertices of phase space, where two of the Mandelstams vanish.

By taking limits of the likelihood, we can establish the boundary behavior of the likelihood.  First, in the collinear limit with, say, $u\to 0$, the likelihood becomes
\begin{align}
\lim_{u\to 0}{\cal L} = \frac{1}{2}- \frac{27}{38}\frac{st}{M^4}\in\left[ \frac{49}{152},\frac{1}{2}\right]\approx\left[0.322,0.5\right]\,.
\end{align}
Note that there is a strange normalization here; ${\cal L} = 1$ is not in the range.  However, we are ignoring overall normalization factors in the amplitudes, so what is most relevant are the locations of extrema and the size of the range of the likelihood, and not its particular values.

In the limit where the final state gluon becomes soft, $u,t\to 0$, the likelihood becomes
\begin{align}
\lim_{u,t\to 0}{\cal L} =  \frac{1}{2}\,,
\end{align}
which is also its maxima.  Unlike for the same limit of $^1S_0$ production, we note that this soft limit returns a single value of the likelihood.  This is because the $^3S_1$ amplitudes are finite in the soft limit, while the $^1S_0$ amplitudes diverge.  The divergence of a higher-point amplitude is indicative of the existence of a lower-point process with the same quantum numbers.  Crossing the production amplitudes, the decay $^1S_0\to gg$ is allowed, but there is no two-body decay $^3S_1\to gg$ at tree-level, because $^3S_1$ is a massive spin-1 particle, in part due to the Landau-Yang theorem \cite{Landau:1948kw,Yang:1950rg}.\footnote{In a non-Abelian gauge theory, the Landau-Yang theorem only forbids decays of spin-1 color-singlets to two gluons \cite{Pleitez:2015cpa,Beenakker:2015mra,Cacciari:2015ela}, because decay of a color-singlet requires that the gluons are indistinguishable.  Tree-level amplitudes of spin-1 color-octet heavy quark pairs decaying to two gluons are widely known to vanish at tree-level \cite{Cho:1995vh,Tang:1995zp,Ma:1995fd,Tang:1996rm,Petrelli:1997ge,Beenakker:2013mva,Barnreuther:2013qvf}, but are non-zero at loop level \cite{Beenakker:2015mra}.}  Therefore, the production of $^3S_1$ from two-gluon scattering is only non-zero if the quarkonium is produced in association with another gluon.  Note, however, that the production process $q\bar q\to \, ^3S_1$ has non-zero amplitude.

Finally, if all of the Mandelstams are identical, then the likelihood takes the value
\begin{align}
\lim_{s=t=u}{\cal L} = \frac{5}{19}\approx 0.263\,,
\end{align}
which is also the point at which it takes its minima.

\subsubsection{A Simple Discrimination Observable}\label{sec:simpdisc}

\begin{figure}[t]
\begin{center}
\includegraphics[width=0.5\textwidth]{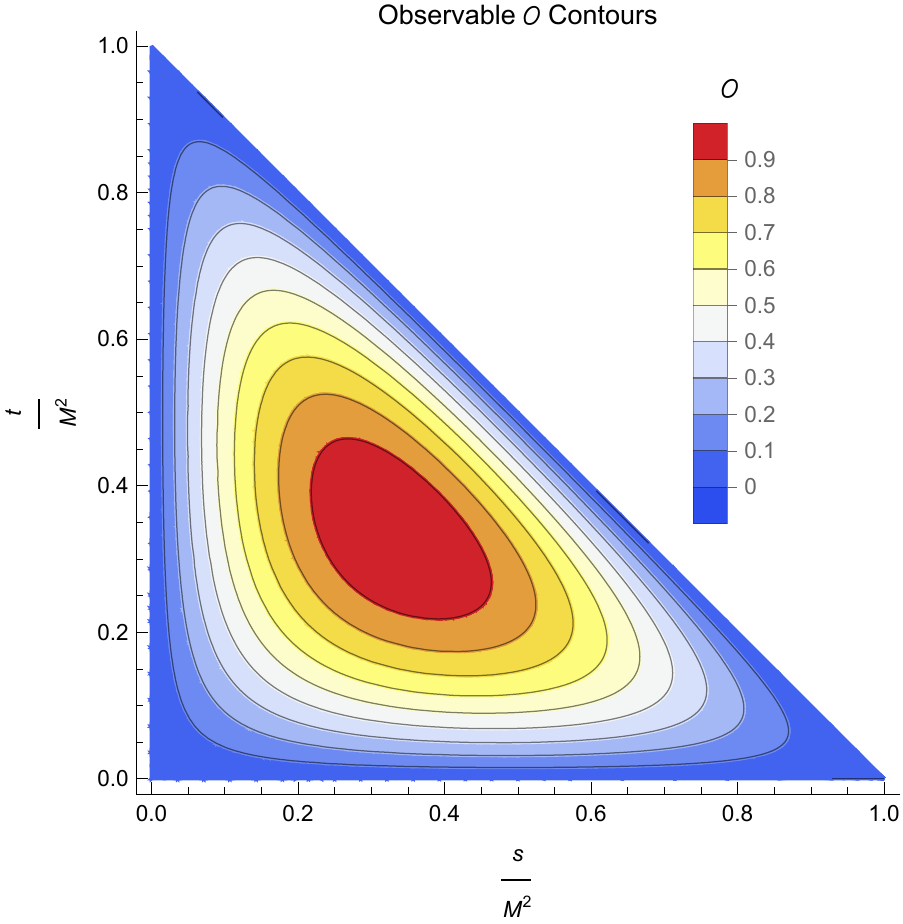}
\caption{\label{fig:oobsplot}
Plot of the contours of observable ${\cal O}$ from \Eq{eq:oobsdef} on $(s,t)$ phase space.}
\end{center}
\end{figure}

We will construct the likelihood ratios from the perturbative amplitudes for $P$-wave heavy quark pair production shortly, but at this point it is useful to pause and note general features of the likelihood ratios for $S$-wave production.  From the functional forms of \Eqs{eq:like1s0}{eq:like3s1}, it is obvious that the likelihood ratios are not identical, nor are they related by a monotonic function of one another.  Because of this, there cannot be a single function for both $^1S_0$ and $^3S_1$ production that is an optimal color-singlet versus -octet discriminant observable on the $(s,t)$ phase space.  Nevertheless, we can construct observables that are good discriminants for both $^1S_0$ and $^3S_1$ production that interpolate between the extrema of the likelihood ratios.  Importantly, the maxima and minima of the two likelihood ratios of \Eqs{eq:like1s0}{eq:like3s1} lie at the same points on $(s,t)$ phase space.  A simple function whose extrema agree with these likelihood ratios is
\begin{align}\label{eq:oobsdef}
{\cal O}\equiv \frac{27stu}{M^6}\,.
\end{align}
This observable is normalized so that it vanishes along all phase space boundaries and takes the value ${\cal O} = 1$ where $s=t=u=M^2/3$.  Contours of this observable on the $(s,t)$ phase space are plotted in \Fig{fig:oobsplot}.

This product observable form for discrimination of color-singlet and color-octet heavy quark pair production is simply related to other observables.  For quarkonium production at a hadron collider, we can express the product $stu$ in natural coordinates.  The colliding partonic gluon momenta are
\begin{align}
&p_1 = x_1 \frac{E_\text{cm}}{2}(1,0,0,1)\,, &p_2 = x_2 \frac{E_\text{cm}}{2}(1,0,0,-1)\,,
\end{align}
where $E_\text{cm}$ is the hadronic center-of-mass collision energy, and $x_1,x_2$ are the momentum fractions of the gluons.  The momenta of the final state gluon and quarkonium can be expressed as
\begin{align}
p_3 &=p_\perp(\cosh\eta,1,0,\sinh\eta)\,,\\
p_{[Q\bar Q]} &= \left(
\sqrt{p_\perp^2+M^2}\cosh y,-p_\perp,0,\sqrt{p_\perp^2+M^2}\sinh y
\right)\,,\nonumber
\end{align}
where $p_\perp$ is their transverse momentum, $\eta$ is the pseudrapidity of the gluon and $y$ is the rapidity of the quarkonium.  In these coordinates, note that
\begin{align}
stu &= s^2 p_\perp^2\\
&=p_\perp^6\left(\frac{M^2}{p_\perp^2}+2+2\sqrt{1+\frac{M^2}{p_\perp^2}}\cosh(y-\eta)
\right)^2\,.\nonumber
\end{align}
Then, the observable ${\cal O}$ can be expressed as
\begin{align}
{\cal O} &=\frac{27stu}{M^6}\\
&= \frac{p_\perp^6}{M^6}\left(\frac{M^2}{p_\perp^2}+2+2\sqrt{1+\frac{M^2}{p_\perp^2}}\cosh(y-\eta)
\right)^2\,.\nonumber
\end{align}

In these collider coordinates, the likelihood ratio for the production of the angular momentum state $^1S_0$ of heavy quark pair from \Eq{eq:like1s0} is
\begin{align}
{\cal L} &=1-\frac{M^2stu}{(st+tu+us)^2}\\
&=1-\frac{M^2}{p_\perp^2\left( 1+2\sqrt{1+\frac{M^2}{p_\perp^2}}\cosh(y-\eta)\right)^2}\,.\nonumber
\end{align}
In the high-boost limit, $p_\perp\to\infty$, the likelihood ratio becomes 1, which is also its maximum.  Correspondingly, the observable ${\cal O}$ in this same high-boost limit diverges, which is its necessarily its maximum.  In the threshold limit, where $p_\perp\to 0$ so that $s\to M^2$, the likelihood ratio reduces to
\begin{align}
\lim_{p_\perp\to 0}{\cal L} \to 1-\frac{1}{4\cosh^2(y-\eta)}\,,
\end{align}
which takes a maximum value of $3/4$ when the rapidity of the quarkonium is the same as the pseudorapidity of the soft final-state gluon.  In this same threshold limit, the observable ${\cal O}$ vanishes, which is necessarily its minimum.  Thus, for production of $S$-wave heavy quark pair, an even simpler observable that extrapolates between the extrema of the likelihood ratio is the transverse momentum of the quarkonium itself.

By crossing all gluons to the final state and the quarkonium to the initial state, then we are considering inclusive hadronic decay of the quarkonium, $[Q\bar Q]\to ggg$.  The decay amplitudes are identical to the production amplitudes, so the observable ${\cal O}$ would also be a powerful discriminant for identification of the color representation of the quarkonium decay products.  The observable ${\cal O}$ can be equivalently expressed as
\begin{align}
{\cal O}&=\frac{27stu}{M^6} \\
&=\frac{27 E_1^2E_2^2E_3^2(1-\cos\theta_{12})(1-\cos\theta_{13})(1-\cos\theta_{23})}{\left(\sum_{1\leq i<j\leq 3}E_iE_j(1-\cos\theta_{ij})\right)^3}\,.
\nonumber
\end{align}
Here, we have labeled the final state gluons as particles $1,2,3$, with $E_i$ the energy of the $i$th gluon and $\theta_{ij}$ the angle between gluons $i$ and $j$.  In the high-boost limit, the cosine factors can be Taylor expanded to
\begin{align}
{\cal O}&\to \frac{27 E_1^2E_2^2E_3^2\theta_{12}^2\theta_{13}^2\theta_{23}^2}{\left(\sum_{1\leq i<j\leq 3}E_iE_j\theta_{ij}^2\right)^3}\,.
\nonumber
\end{align}
This can be generalized into a ratio of infrared and collinear safe observables that sum over an arbitrary number of decay products as:
\begin{align}
{\cal O}\to 
\frac{36\left(\sum_{i,j,k}E_iE_jE_k\theta_{ij}\theta_{ik}\theta_{jk}\right)^2}{\left(\sum_{i,j}E_i E_j\theta_{ij}^2\right)^3}
\,,
\end{align}
where the sums run over all decay products.  In this form, this observable is essentially identical to the observable $D_2$ \cite{Larkoski:2014gra}, which was constructed in jet substructure studies for identification of multi-pronged high-energy jets.  While not initially constructed for discrimination of distinct color representations of boosted objects, $D_2$ has been demonstrated to be a powerful color-singlet vs.~color-octet discriminant in other scenarios \cite{Buckley:2020kdp}.

For this observable ${\cal O}$ to be a good discriminant for any process for quarkonium production, the likelihood ratios between color-singlet and -octet production for all angular momentum states must have maxima and minima in the same location.  This analysis of $S$-wave states is suggestive of this general property, but we still need to analyze $P$-wave states.  In the following, we will find that most of the $P$-wave states actually have their points of maxima and minima flipped as compared to $S$-wave states, and this will lead to an obstruction for construction of a general discrimination observable.

\subsection{$P$-wave States}

\subsubsection{$^3P_0$}

\begin{figure}[t]
\begin{center}
\includegraphics[width=0.5\textwidth]{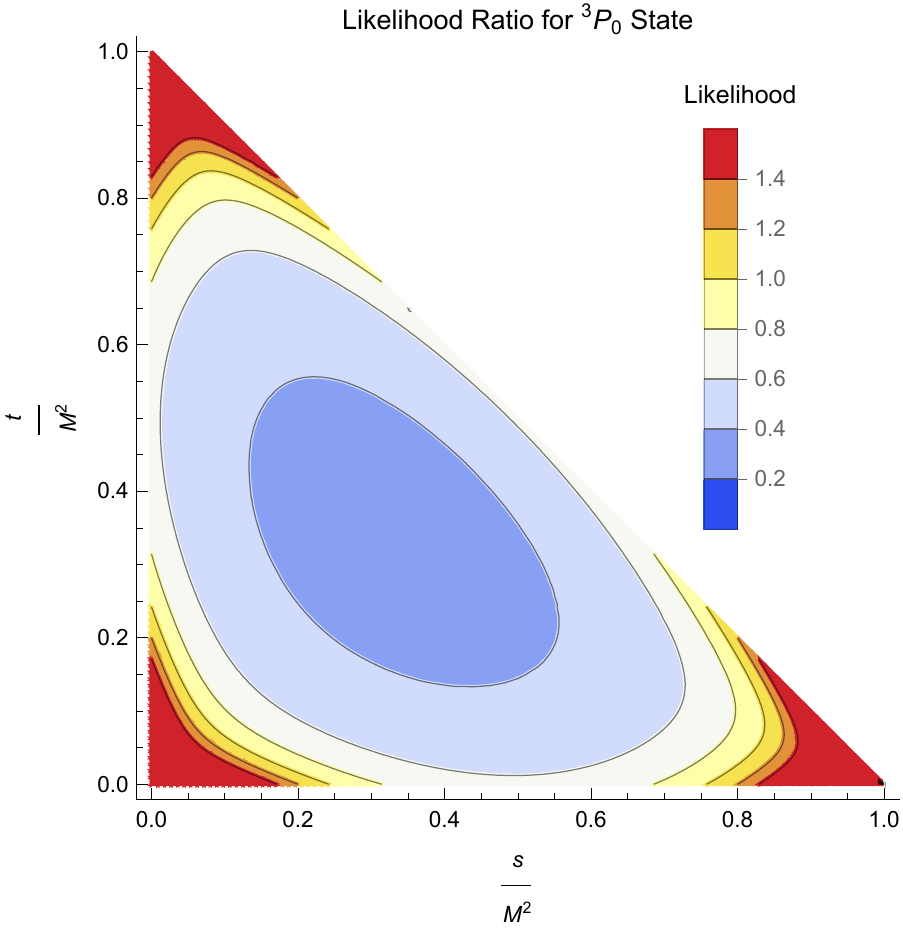}
\caption{\label{fig:like3p0}
Plot of the perturbative octet vs.~singlet likelihood ratio on $(s,t)$ phase space for the $^3P_0$ state.  Contours in the corners, where a gluon becomes soft, have been truncated at ${\cal L} = 1.5$.}
\end{center}
\end{figure}

Starting with the heavy quark pair angular momentum state $^3P_0$, the likelihood ratio between the color-singlet and -octet squared matrix elements is defined as
\begin{align}
{\cal L}=\frac{|{\cal A}(gg\to \,\! ^{3}P_0^{(8)} g)|^2}{|{\cal A}(gg\to \,\! ^{3}P_0^{(1)} g)|^2}\,.
\end{align}
The explicit expressions for the amplitudes as functions of the Mandelstam invariants $s,t,u$ are provided in \App{app:3p0me}.  A contour plot of the likelihood ratio on the $(s,t)$ phase space is shown in \Fig{fig:like3p0}.  As observed for the $S$-wave states, the likelihood ratio takes a minimum value at the center of phase space, where $s=t=u=M^2/3$, and a maximum on the boundaries of phase space.

The value of the likelihood in these limits takes simple forms.  First, in the collinear limit, in which $u\to 0$, the likelihood becomes
\begin{align}
\lim_{u\to 0}{\cal L} = \frac{(s^2+st+t^2)^2}{2st(2s^2+3st+2t^2)}\in\left[ \frac{9}{14},\infty\right)\,.
\end{align}
Further, in the limit where a gluon becomes soft, the likelihood diverges
\begin{align}
\lim_{u,t\to 0}{\cal L} \to \infty\,.
\end{align}
Unlike observed for the $S$-wave states, the color-singlet $^3P_0$ state is arbitrarily smaller than the corresponding color-octet state, in the soft gluon limit.  That means that, for this angular momentum state, a heavy quark pair associated with a soft gluon is necessarily in a color-octet state, with no color-singlet contamination.

At the point where all of the Mandelstam invariants are equal, $s=t=u=M^2/3$, the likelihood takes the value
\begin{align}
\lim_{s=t=u}{\cal L} = \frac{127}{369}\approx 0.344\,,
\end{align}
which is also its minimum on phase space.

\subsubsection{$^3P_1$}

\begin{figure}[t]
\begin{center}
\includegraphics[width=0.5\textwidth]{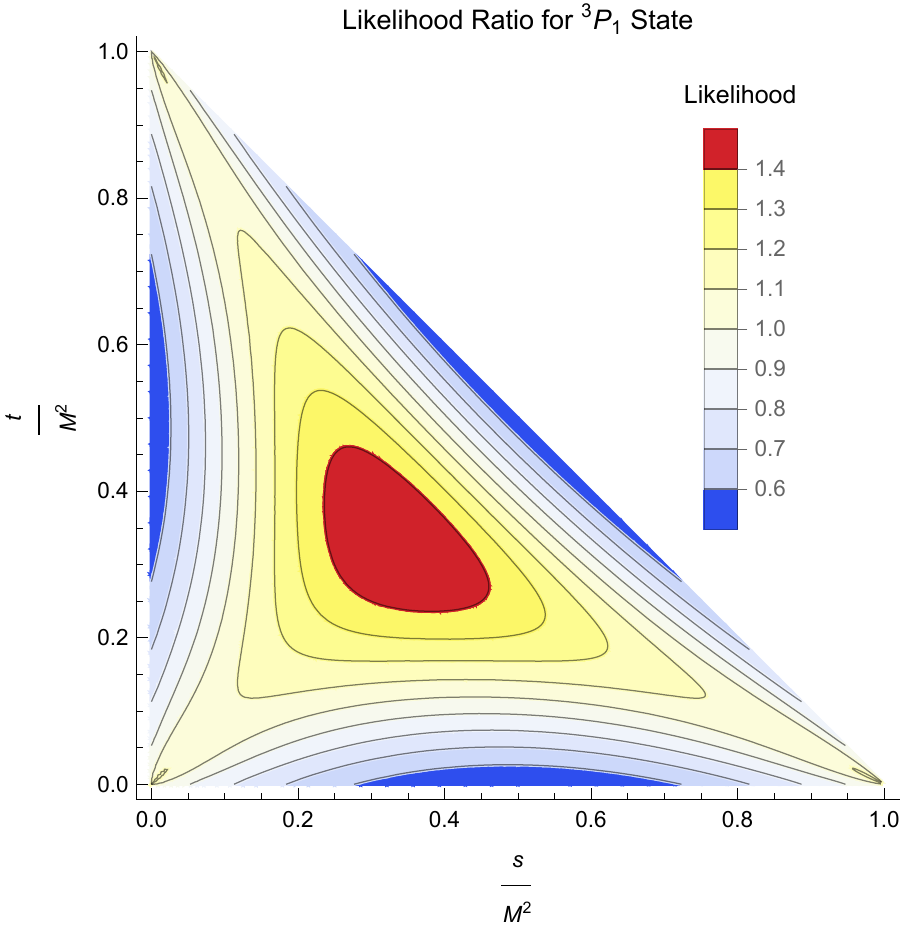}
\caption{\label{fig:like3p1}
Plot of the perturbative octet vs.~singlet likelihood ratio on $(s,t)$ phase space for the $^3P_1$ state.}
\end{center}
\end{figure}

The likelihood ratio between the color-singlet and color-octet squared matrix elements is defined as
\begin{align}
{\cal L}=\frac{|{\cal A}(gg\to \,\! ^{3}P_1^{(8)} g)|^2}{|{\cal A}(gg\to \,\! ^{3}P_1^{(1)} g)|^2}\,.
\end{align}
The explicit expressions for the amplitudes as functions of the Mandelstam invariants $s,t,u$ are provided in \App{app:3p1me}.  A contour plot of the likelihood ratio on the $(s,t)$ phase space is shown in \Fig{fig:like3p1}.  Unlike any angular momentum states considered before, the likelihood now has a maximum at the center of phase space, and a minimum on the edges.  Extrema still lie at the same locations as all other heavy quark pair states considered before, and so the observable ${\cal O}$ introduced in \Sec{sec:simpdisc} is still a good color-singlet vs.~-octet discriminant on this state.  However, depending on the angular momentum states that contribute for the quarkonium of interest, the extrema of the relevant likelihood ratio is not guaranteed to lie at the center and along the edges of phase space.  So, the form of the discrimination observable as a function of the Mandelstam invariants would have non-trivial dependence on angular momentum and the long-distance matrix elements, severely limiting a general analysis applicable to any quarkonium state.

To understand the extrema of this likelihood ratio more, it is useful to take its limits.  In the collinear limit, $u\to 0$, the likelihood ratio simplifies to
\begin{align}
\lim_{u\to 0}{\cal L} = \frac{s^2+t^2}{(s+t)^2}\in\left[\frac{1}{2},1\right]\,.
\end{align}
In the soft limit, $t,u\to 0$, the likelihood becomes
\begin{align}
\lim_{t,u\to 0}{\cal L} \to \frac{(t^2+tu+u^2)^2}{(t+u)^2(t^2+u^2)}\in\left[
1,\frac{9}{8}
\right] \,.
\end{align}
Finally, in the hard limit where $s=t=u=M^2/3$, the likelihood becomes
\begin{align}
\lim_{s=t=u}{\cal L} = \frac{3}{2}\,.
\end{align}

\subsubsection{$^3P_2$}

\begin{figure}[t!]
\begin{center}
\includegraphics[width=0.5\textwidth]{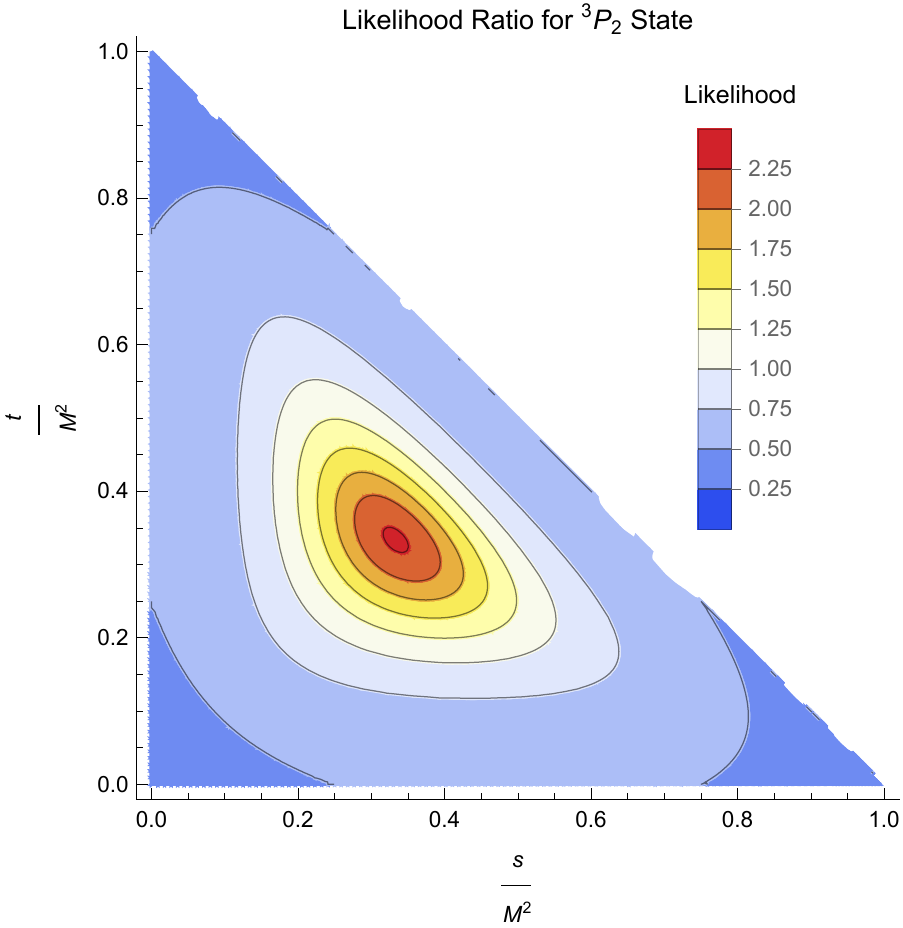}
\caption{\label{fig:like3p2}
Plot of the perturbative octet vs.~singlet likelihood ratio on $(s,t)$ phase space for the $^3P_2$ state.}
\end{center}
\end{figure}

The likelihood ratio for the color-singlet vs.~-octet $^3P_2$ heavy quark pair state exhibits similar extrema as the $^3P_1$ state.  The likelihood ratio is
\begin{align}
{\cal L}=\frac{|{\cal A}(gg\to \,\! ^{3}P_2^{(8)} g)|^2}{|{\cal A}(gg\to \,\! ^{3}P_2^{(1)} g)|^2}\,,
\end{align}
and explicit expressions for the amplitudes as functions of the Mandelstam invariants $s,t,u$ are provided in \App{app:3p2me}.  A contour plot of the likelihood ratio on the $(s,t)$ phase space is shown in \Fig{fig:like3p2}, from which the maximum is at the center of phase space, and the minima on its edges.  In the collinear limit, $u\to 0$, the likelihood ratio simplifies to
\begin{align}
\lim_{u\to 0}{\cal L} = \frac{1}{2}\,.
\end{align}
In the soft limit, $t,u\to 0$, the likelihood becomes
\begin{align}
\lim_{t,u\to 0}{\cal L} \to \frac{t^2+tu+u^2}{2(t+u)^2}\in\left[
\frac{3}{8},\frac{1}{2}
\right] \,.
\end{align}
Finally, in the hard limit where $s=t=u=M^2/3$, the likelihood becomes
\begin{align}
\lim_{s=t=u}{\cal L} = \frac{41}{18}\simeq 2.28\,.
\end{align}

\section{Conclusions}\label{sec:conc}

In this paper, we compared the lowest-order, short-distance, perturbative amplitudes for heavy quark pair production at a hadron collider.  For fixed angular momentum, we demonstrated that the likelihood ratio of color-octet to color-singlet heavy quark pair production amplitudes has a structure on phase space that depends on the angular momentum.  For a fixed angular momentum state, a discrimination observable that interpolates between the extrema of the likelihood is simple to construct.  However, in a more realistic experimental scenario, cross sections of multiple heavy quark pair states will need to be summed together consistent with the quantum numbers of the observed hadron, weighted by LDMEs.  The dependence of the location of maxima and minima of the likelihood ratios of individual angular momentum states means that, generically, a robust color-singlet versus color-octet production discrimination observable cannot be so easily constructed.  This also means that regions of phase space where color-singlet or color-octet production are purified, or one becomes dominant, are challenging to determine for quarkonia in general.  

However, this is not a no-go theorem that would forbid any discrimination observable to be constructed.  A proposed discrimination observable would depend on the hadron of interest, and could be studied in particular cases, like for $\chi_J$ production.  Within NRQCD, the lowest-order states in the velocity expansion that contribute to inclusive $\chi_J$ production result in a cross section of the form
\begin{align}
d\sigma_{\chi_J} = d\hat \sigma_{^3P_J}^{(1)}\langle {\cal O}_{\,^3P_J}^{(1)}\rangle+d\hat \sigma_{^3S_1}^{(8)}\langle {\cal O}_{\,^3S_1}^{(8)}\rangle +\cdots\,.
\end{align}
The color-singlet versus color-octet likelihood ratio is then
\begin{align}
{\cal L} = \frac{d\hat \sigma_{^3S_1}^{(8)}\langle {\cal O}_{\,^3S_1}^{(8)}\rangle}{d\hat \sigma_{^3P_J}^{(1)}\langle {\cal O}_{\,^3P_J}^{(1)}\rangle}\,.
\end{align}
For each value of total angular momentum $J$, this likelihood ratio takes a distinct functional form, according to the short-distance matrix elements.  However, with the relevant leading-order short-distance matrix elements provided here, we demonstrate in \App{app:jpsi} that the functional structure of this likelihood on phase space is highly sensitive to the total angular momentum.  A more detailed study with experimentally-extracted values for the LDMEs for specific, observed hadrons would be interesting to possibly identify if there are any general features to the color-singlet vs.~color-octet heavy quark pair production likelihood ratio.

In this direction of a more realistic study, one could include next-to-leading order perturbative corrections to the short-distance matrix elements.  The analytic matrix element analysis performed here would not be sustainable when higher orders are included, and at any rate, infrared divergences would need to be inclusively summed over for finite predictions.  This would correspondingly require a phase space definition with a collection of infrared and collinear safe observables, along the lines of that proposed in \Ref{Datta:2017rhs}.  There exist some numerical codes for higher-order predictions of quarkonium production, such as \Ref{Wan:2014vka} and \Ref{Brambilla:2020fla}.  However, the desired highly-differential nature of the likelihood on phase space may be very challenging to achieve at next-to-leading order, due to the curse of dimensionality and the difficulty of real and virtual divergence cancellation.  Nevertheless, higher-order predictions would establish the robustness of results presented here.

Along exclusive phase space boundaries, a mismatch between real and virtual contributions generally produces large logarithms of the distance to the boundary, and spoils fixed-order perturbative convergence.  To restore predictive power, one must resum these large logarithms to all orders in the coupling, which typically results in exponential suppression of emissions near the boundary.  Including logarithmic resummation into the short-distance matrix elements can be accomplished analytically or through a numerical parton shower, like the interface of Pythia 8 \cite{Sjostrand:2007gs} with HELAC-Onia \cite{Shao:2015vga}.  Because we observed that several of the specific angular momentum state likelihood ratios exhibited maxima at the boundaries of phase space, resummation will likely push the maxima away from the boundaries, to the interior of phase space.  This effect may correspondingly result in even greater sensitivity of the likelihood to the specific angular momentum state.  Nevertheless, more detailed and focused studies are clearly required to conclusively answer the question of the mechanism of quarkonium production.

\begin{acknowledgments}

I thank Pierre Artoisenet for discussions long ago about color-singlet and color-octet discrimination observables.  I thank Tom Mehen and Adam Leibovich for comments and clarifications of predictions of NRQCD.  This work was supported by the Department of Energy, Contract DE-AC02-76SF00515.

\end{acknowledgments}

\appendix

\begin{widetext}

\section{Matrix Elements}\label{app:me}

\subsection{$^3P_0$ Matrix Elements}\label{app:3p0me}

The squared color-singlet matrix element is \cite{Gastmans:1987be}
\begin{align}
|{\cal A}(gg\to \,\! ^{3}P_0^{(1)} g)|^2 &\propto \frac{1}{(s+t)^2(t+u)^2(u+s)^2}\\
&
\hspace{1cm}
\times \left[
8M^2\left(
\frac{tu(t^4-t^2u^2+u^4)}{(t+u)^2}+\frac{us(u^4-u^2s^2+s^4)}{(u+s)^2}+\frac{st(s^4-s^2t^2+t^4)}{(s+t)^2}
\right)\right.\nonumber\\
&
\hspace{2cm}
+4M^4\left(
M^2(st+tu+us)-5stu
\right)+\frac{9M^8}{stu}(st+tu+us)^2\nonumber\\
&
\hspace{2cm}
+\frac{(st+tu+us)^2}{(s+t)(t+u)(u+s)}\left(
\phantom{\frac{1}{s}}
\!\!\!\!
8M^4(s^2+t^2+u^2)-16M^2stu\right.\nonumber\\
&
\hspace{2cm}
\left.\left.
+\left(
1-9M^2\left(
\frac{1}{s}+\frac{1}{t}+\frac{1}{u}
\right)
\right)(s^4+t^4+u^4)
\right)
\right]\,.\nonumber
\end{align}
The squared color-octet matrix element is \cite{Cho:1995vh,Cho:1995ce}
\begin{align}
|{\cal A}(gg\to \,\! ^{3}P_0^{(8)} g)|^2 &\propto \frac{1}{stu(s+t)^4(t+u)^4(u+s)^4}\left[
2 s^4 t^4 u^4 \left(3676 s^2+4683 s t\right)\phantom{\left(s^2\right)^2}\right.\\
&
\hspace{1cm}+s^3 t^3 u^3
   \left(1336 s^5+3533 \left(s^4 t+s t^4\right)+5720
   \left(s^3 t^2+s^2 t^3\right)\right)\nonumber\\
   &
   \hspace{1cm}+s^2 t^2 u^2
   \left(36 s^8+288 \left(s^7 t+s t^7\right)+2624 s^4
   t^4+1009 \left(s^6 t^2+s^2 t^6\right)+2067 \left(s^5
   t^3+s^3 t^5\right)\right)\nonumber\\
   &
   \hspace{1cm}+3 s t^3 u^3 (t+u) \left(9
   t^6+43 t^5 u+94 t^4 u^2+124 t^3 u^3+94 t^2 u^4+43 t
   u^5+9 u^6\right)\nonumber\\
   &
   \hspace{1cm}\left.+\,9 t^4 u^4 (t+u)^2 \left(t^2+t
   u+u^2\right)^2+(stu)\to(tus)+(stu)\to (ust)
\right]\,.\nonumber
\end{align}

\subsection{$^3P_1$ Matrix Elements}\label{app:3p1me}

The squared color-singlet matrix element is \cite{Gastmans:1987be}
\begin{align}
|{\cal A}(gg\to \,\! ^{3}P_1^{(1)} g)|^2 &\propto \frac{1}{(s+t)^2 (s+u)^2 (t+u)^2}\\
&\hspace{1cm}
\times
\left[\frac{2 \left(s^2 t^2+s^2 u^2+t^2 u^2\right) \left(M^2 s t u+s^2 t^2+s^2 u^2+t^2
   u^2\right)}{(s+t) (s+u) (t+u)}\right.\nonumber\\
   &
   \hspace{2cm}\left.+M^2 \left(\frac{s^2 t^2
   \left(s^2+t^2\right)}{(s+t)^2}+\frac{s^2 u^2
   \left(s^2+u^2\right)}{(s+u)^2}+\frac{t^2 u^2
   \left(t^2+u^2\right)}{(t+u)^2}\right)\right]\,.\nonumber
\end{align}
The squared color-octet matrix element, when summed over constituent quark spins, is \cite{Cho:1995vh,Cho:1995ce}
\begin{align}
|{\cal A}(gg\to \,\! ^{3}P_1^{(8)} g)|^2 &\propto \frac{1}{(s+t)^4 (s+u)^4 (t+u)^4}\left[
s^3 t^3 u^3 \left(57 s^2+83 s t\right)\right.\\
&
\hspace{1cm}
+s^2 t^2 u^2 \left(3 s^5+17 s^4 t+38 s^3 t^2+38 s^2
   t^3+17 s t^4\right)+s t^3 u^3 \left(2 t^4+8 t^3 u+11 t^2 u^2+8 t u^3+2 u^4\right)\nonumber\\
   &\hspace{1cm}\left.+\,t^4
   u^4 (t+u) \left(t^2+u^2\right)+(stu)\to(tus)+(stu)\to (ust)
\right]\,.\nonumber
\end{align}

\subsection{$^3P_2$ Matrix Elements}\label{app:3p2me}

The squared color-singlet matrix element is \cite{Gastmans:1987be}
\begin{align}
|{\cal A}(gg\to \,\! ^{3}P_2^{(1)} g)|^2 &\propto \frac{1}{(s+t)^2 (s+u)^2 (t+u)^2}\\
&
\hspace{1cm}
\times \left[
M^2 \left(\frac{s^2 t^2 \left(s^2+4 s t+t^2\right)}{(s+t)^2}+\frac{s^2 u^2 \left(s^2+4
   s u+u^2\right)}{(s+u)^2}+\frac{t^2 u^2 \left(t^2+4 t u+u^2\right)}{(t+u)^2}\right)\right.\nonumber\\
   &
   \hspace{2cm}
+   12 M^2 \left(3 \left(s^3 t+s^3 u+s t^3+s u^3+t^3 u+t u^3\right)+4 M^2 s t u\right)\nonumber\\
&
\hspace{2cm}
+\frac{2 (s t+s u+t u)^2 \left(s t+s u+t u-M^4\right)}{(s+t) (s+u) (t+u)} \left(s t+s u+t u-24
   M^4\phantom{\frac{1}{s}}\right.\nonumber\\
   &\hspace{3cm}
   \left.\left.-6 M^2
   \left(\frac{1}{s}+\frac{1}{t}+\frac{1}{u}\right) \left(s t+s u+t u-M^4\right)\right)
\right]\,.\nonumber
\end{align}
The squared color-octet matrix element, when summed over constituent quark spins, is \cite{Cho:1995vh,Cho:1995ce}
\begin{align}
|{\cal A}(gg\to \,\! ^{3}P_2^{(8)} g)|^2 &\propto \frac{1}{stu(s+t)^4 (s+u)^4 (t+u)^4}\left[
s^4 t^4 u^4 \left(2555 s^2+2937 s t\right)\phantom{\left(s^2\right)^2}\right.\\
&
\hspace{1cm}+s^3 t^3 u^3 \left(751 s^5+1556 s^4 t+2096 s^3
   t^2+2096 s^2 t^3+1556 s t^4\right)\nonumber\\
   &\hspace{1cm}+s^2 t^2 u^2 \left(24 s^8+189 s^7 t+562 s^6 t^2+951
   s^5 t^3+1109 s^4 t^4+951 s^3 t^5+562 s^2 t^6+189 s t^7\right)\nonumber\\
   &
   \hspace{1cm}+3 s t^3 u^3 (t+u)
   \left(6 t^6+27 t^5 u+51 t^4 u^2+62 t^3 u^3+51 t^2 u^4+27 t u^5+6 u^6\right)\nonumber\\
   &\hspace{1cm}\left.+\,6 t^4 u^4
   (t+u)^2 \left(t^2+t u+u^2\right)^2+(stu)\to(tus)+(stu)\to (ust)
\right]\,.\nonumber
\end{align}

\end{widetext}

\section{$\chi_J$ Production Likelihood}\label{app:jpsi}

In this appendix, we present plots of color-octet to color-singlet likelihood ratios for $\chi_J$ production in NRQCD, to leading order in the velocity expansion.  As presented in the conclusions, the likelihood ratio in NRQCD is
\begin{align}
{\cal L} = \frac{d\hat \sigma_{^3S_1}^{(8)}\langle {\cal O}_{\,^3S_1}^{(8)}\rangle}{d\hat \sigma_{^3P_J}^{(1)}\langle {\cal O}_{\,^3P_J}^{(1)}\rangle}\,,
\end{align}
which is some function of the ratio of short-distance squared amplitudes for heavy quark pair production, for each value of total angular momentum $J$.  By plotting these ratios from the leading-order amplitudes provided above, we show that their structure on phase space depends on the value of $J$, obfuscating the construction of a single color-octet versus color-singlet  discrimination observable for any $J$.

\begin{figure}[t!]
\begin{center}
\includegraphics[width=0.5\textwidth]{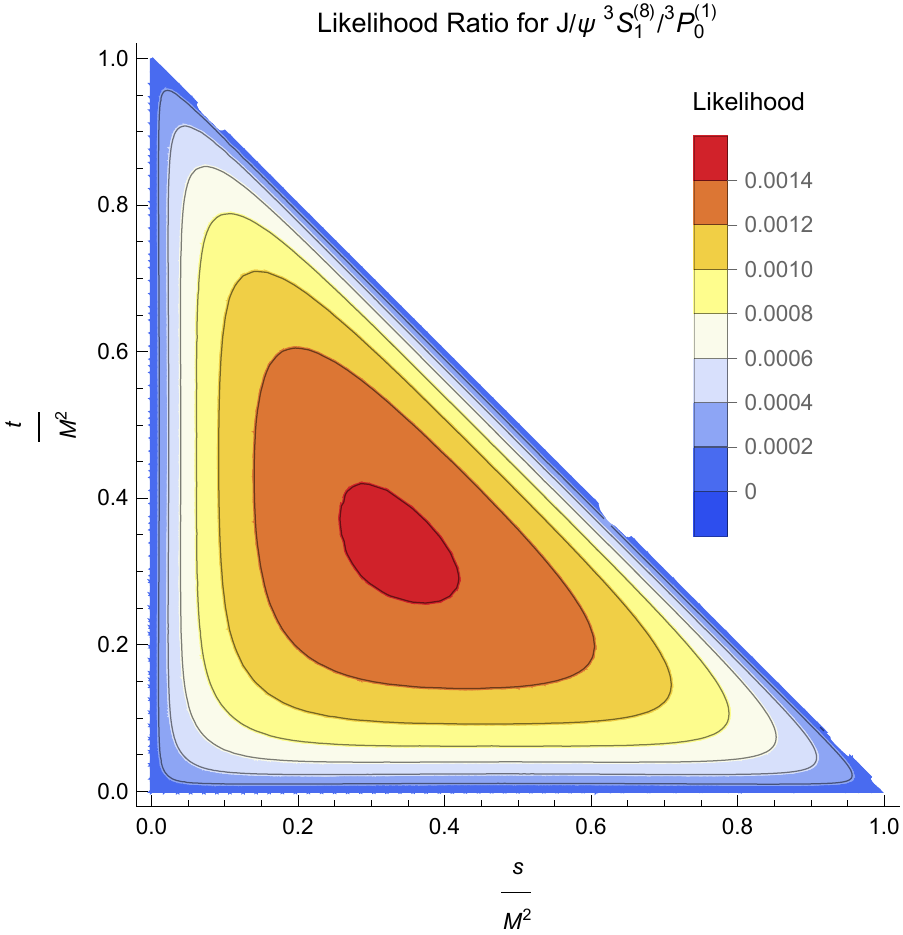}
\caption{\label{fig:likejpsi3p0}
Plot of the perturbative likelihood ratio of the $^3S_1^{(8)}$  to $^3P_0^{(1)}$ states on $(s,t)$ phase space.}
\end{center}
\end{figure}

\begin{figure}[t!]
\begin{center}
\includegraphics[width=0.5\textwidth]{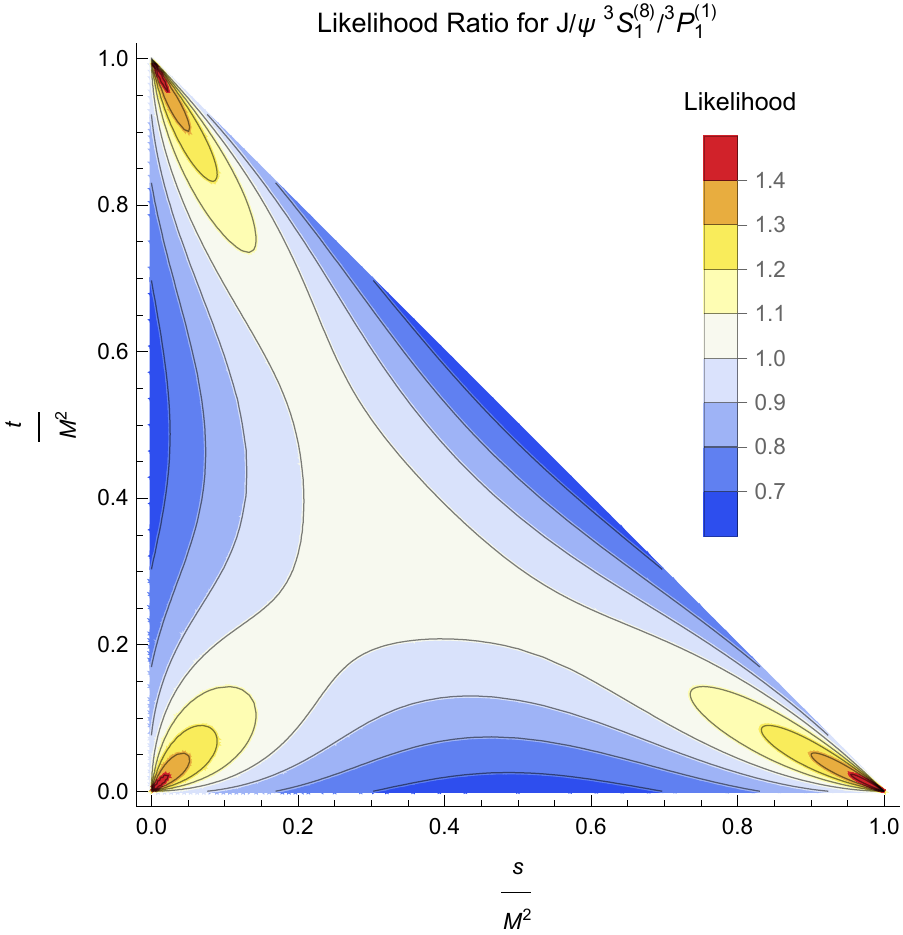}
\caption{\label{fig:likejpsi3p1}
Plot of the perturbative likelihood ratio of the $^3S_1^{(8)}$  to $^3P_1^{(1)}$ states on $(s,t)$ phase space.}
\end{center}
\end{figure}

\begin{figure}[t!]
\begin{center}
\includegraphics[width=0.5\textwidth]{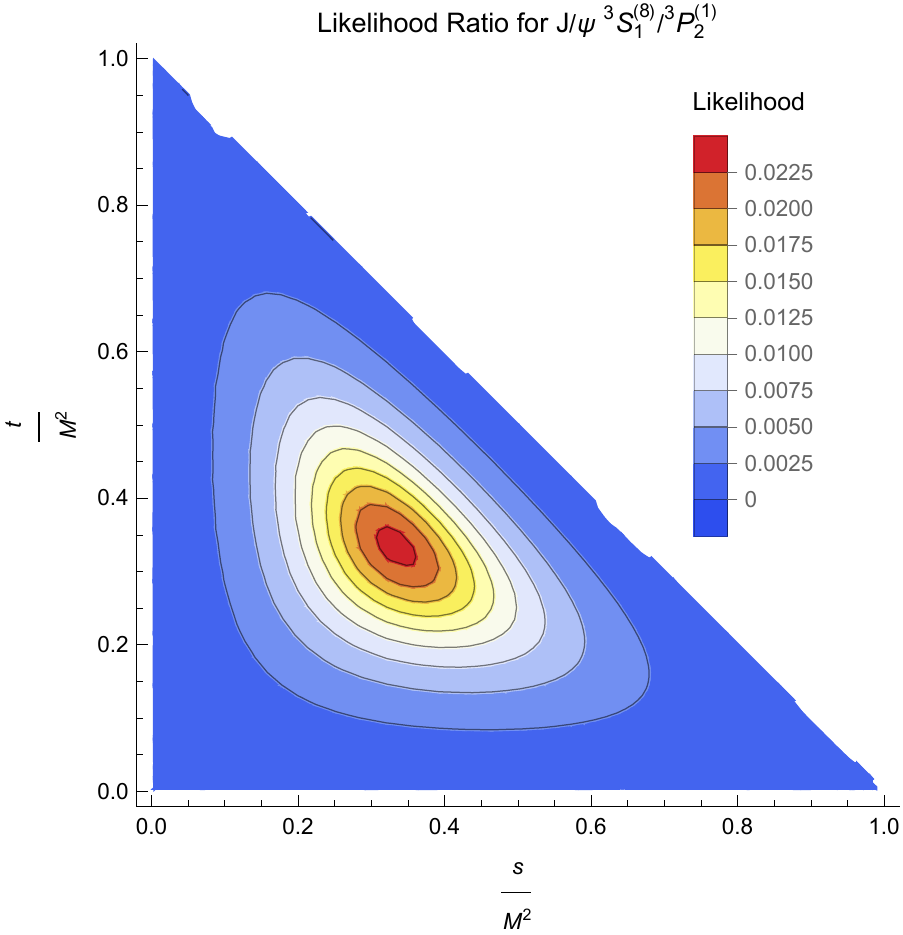}
\caption{\label{fig:likejpsi3p2}
Plot of the perturbative likelihood ratio of the $^3S_1^{(8)}$  to $^3P_2^{(1)}$ states on $(s,t)$ phase space.}
\end{center}
\end{figure}

In \Fig{fig:likejpsi3p0}, \Fig{fig:likejpsi3p1}, and \Fig{fig:likejpsi3p2}, we plot these perturbative likelihood ratios in the $(s,t)$ phase space for quarkonium decay, for each state $^3P_J^{(1)}$ with value of total angular momentum $J$.  Likelihood ratios involving the $^{3}P_0^{(1)}$ and $^{3}P_2^{(1)}$ states exhibit minima and maxima at the same points on phase space, namely at the boundaries and the center, respectively.  However, the likelihood with the $^{3}P_1^{(1)}$ state is largest at the vertices of phase space, and has a saddle point at the center of phase space.

\bibliography{quarkonia}

\end{document}